\documentclass[]{mn2e}
\usepackage{graphicx}

\def\98a{SN~1998A}
\def\87a{SN~1987A}

\def\kms{km s$^{-1}$}

\def\m100{mag/100$^d$}

\def\c57{{$^{57}$Co}\/}

\def\ti44{{$^{44}$Ti}\/}

\begin{document}

\title[SN 1998A: Explosion of a BSG] { SN
1998A: Explosion of a Blue Supergiant } \author[Pastorello
et al. ] {A. Pastorello$^{1,2}$
\thanks{e-mail: pasto@MPA-Garching.MPG.DE}, 
E. Baron$^{3}$, D. Branch$^{3}$, L. Zampieri$^{2}$,\and
 M. Turatto$^{2}$, M. Ramina$^{2}$, S. Benetti$^{2}$, E. Cappellaro$^{4}$, M. Salvo$^{5}$,\and
 F. Patat$^{6}$, A. Piemonte $^{6}$, J. Sollerman$^{7}$, B. Leibundgut$^{6}$, G. Altavilla$^{8}$\\
 $^{1}$ Max Planck Institut f\"{u}r Astrophysik, Karl-Schwarzschild-Str. 1, D-85741 Garching bei M\"unchen, Germany\\
 $^{2}$ INAF - Osservatorio Astronomico di Padova, Vicolo dell'
Osservatorio 5, I-35122 Padova, Italy\\
 $^{3}$ Department of Physics
and Astronomy, University of Oklahoma, 440 W. Brooke St., Norman, OK
73019, USA\\
 $^{4}$ INAF - Osservatorio Astronomico di Napoli, Via Moiariello 16,
I-80131 Napoli, Italy\\
 $^{5}$ Australian National University, Mt. Stromlo Observatory,
2611 Weston ACT, Australia\\
 $^{6}$ European Southern Observatory,
Karl-Schwarzschild-Str. 2, D-85748, Garching bei M\"unchen, Germany\\
 $^{7}$ Stockholm Observatory, Alba Nova University Center,
Roslagstullsbacken 21, SE-106 91 Stockholm, Sweden\\
 $^{8}$ Department of Astronomy, University of Barcelona, Mart\'{i} i Franqu\'{e}s 1, E-08028 Barcelona, Spain}

 \date{Accepted
.....; Received ....; in original form ....}

\maketitle

\begin{abstract}
  We present spectroscopic and photometric observations of the
  peculiar Type II supernova (SN) 1998A. The light curves and spectra
  closely resemble those of \87a, suggesting that the \98a\/
  progenitor exploded when it was a compact blue supergiant. However,
  the comparison with SN 1987A also highlights some important
  differences: \98a\/ is more luminous and the spectra show bluer
  continua and larger expansion velocities at all epochs.  These
  observational properties indicate that the explosion of \98a\/ is
  more energetic than SN~1987A and more typical of SNe II. Comparing
  the observational data to simulations, we deduce that the progenitor
  of \98a\/ was a massive star ($\sim$ 25 M$_{\odot}$) with a small
  pre--supernova radius ($\la$ 6 $\times$ 10$^{12}$ cm).  The Ba II
  lines, unusually strong in \87a and some faint II--P events, are
  almost normal in the case of \98a, indicating that the temperature
  plays a key role in determining their strength.
\end{abstract}


\section{Introduction}
The explosion of \87a\/ in the Large Magellanic Cloud (LMC) led the
question whether this supernova (SN) was unique or if other similar
supernovae (SNe) had been observed.  A number of publications
speculated on possible similarities with past events [e.g. SN~1909A
\cite{bran89} and SN~1982F \cite{vand89}], but none showed an unequivocal
resemblance with \87a\/.  Astronomers had to wait 11 years after
\87a\/ to identify another event showing a clear observational
analogy:
\98a\/ \cite{woo98a,woo98b}.\\

\begin{figure}
\includegraphics[width=8.5cm]{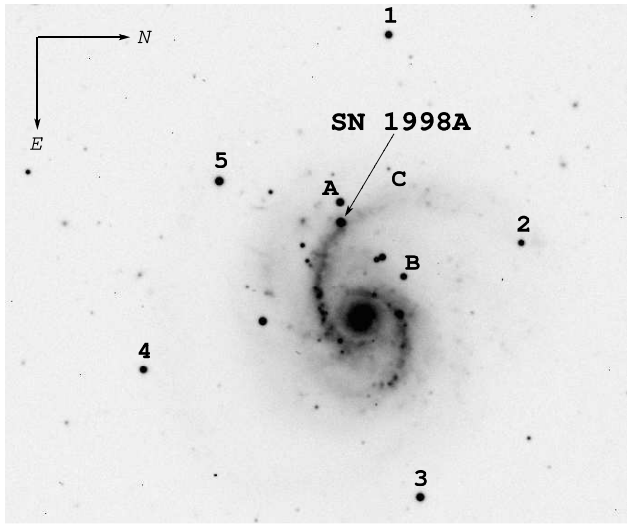}
\caption{SN 1998A and local reference stars in the field of IC
2627 (R filter image, obtained on 1998 March 22 with the Danish
1.54m telescope equipped with DFOSC). The
optical reference stars are labelled with numbers, the infrared ones
with letters.  East is down, North is to the
right. \label{field98A}} 
\end{figure}

\98a\/ was discovered with the 0.61m Perth--Lowell reflector, 
during the Automated Supernova Search Program of the Perth 
Astronomy Research Group (PARG) on 1998 January 6.77 UT \cite{will98}.  
The position of the SN was $\alpha$ = 11$^{h}$09$^{m}$50$\fs$33, $\delta$ =
$-23\degr43\arcmin43\farcs1$ (equinox 2000.0), in the southern
arm of the SBc galaxy IC 2627, 40$^{\prime\prime}$ West and 10$^{\prime\prime}$ South
\cite{will98,fili98} from the nucleus (see Fig. \ref{field98A}).
The discovery magnitude was R $\sim$ 17, and a prediscovery limit (m $>$ 18.5) 
obtained on 1997 December 18 was reported by Williams et
al. \shortcite{will98}.  Filippenko $\&$ Moran (1998) classified \98a\/
as a spectroscopically normal Type II event with strong P--Cygni features.  But about 70
days after the discovery, Woodings et al. \shortcite{woo98a} noted
that the luminosity of \98a\/ has increased by about 1 magnitude,
reaching a broad maximum. They concluded that this photometric
behaviour was analogous to that observed in \87a.  Later on, the PARG group
published VRI photometry of \98a\/ covering up to  $\sim$ 160 days
after discovery \cite{woo98b}. Their photometry confirmed the
similarity with the behaviour of \87a.

The basic informations of \98a\/ and its host galaxy IC 2627 are given in
Tab. \ref{hostgal}.

We present well sampled photometry and spectroscopy
of \98a\/~covering a period of more than 1 year after the explosion and obtained
with several telescopes at ESO--La Silla. In Sect. 2 we present 
the observations and briefly describe the steps of
data reduction.
In Sect. 3 we present the photometric data and a comparison between
the light and colour curves of \98a\/ and
\87a. In Sect. 4 we show the spectroscopic data, discuss the
analogies with spectra of \87a and compare the earliest
spectrum of \98a\/ with synthetic spectra computed with the codes {\sl SYNOW} \cite{fish00}
and \texttt{PHOENIX} \cite{hbjcam99}. A discussion follows in
Sect. 5, where the data of \98a\/ are compared to simulations. The
implications of observational 
similarities between \98a\/ and \87a\/ are analysed, with particular
focus on the nature of the progenitor stars.  A short summary follows in Sect. 6. 

\begin{table}
\caption{Main data of \98a~and the host galaxy IC 2627.} \label{hostgal}
\begin{center}
\begin{tabular}{|c|c|c|} \\ \hline
\multicolumn{3}{|c|}{SN 1998A} \\ \hline
$\alpha$ (J2000.0) & 11$^{h}$09$^{m}$50\fs33 & $\otimes$ \\
$\delta$ (J2000.0) & -23$\degr$43$\arcmin$43$\farcs$.1 & $\otimes$ \\
Offset SN - Gal. Nucleus & 40$^{\prime\prime}$W, 10$^{\prime\prime}$S & $\otimes$, $\diamond$ \\
SN Type & IIpec & $\diamond$, $\circ$ \\
Discovery Date (UT)& 1998 January 6.77 & $\otimes$ \\
Discovery Julian Date & 2450820.27 & $\otimes$ \\
Assumed explosion JD & 2450801 & $\circ$ \\
Discovery R Magnitude & 17 & $\otimes$ \\ \hline
\multicolumn{3}{|c|}{IC 2627} \\ \hline
$\alpha$ (J2000.0) & 11$^{h}$09$^{m}$53$\fs$.46 & $\odot$ \\
$\delta$ (J2000.0) & --23$\degr$43$\arcmin$35$\arcsec$.3 & $\odot$\\
Morphological Type & SBc & $\nabla$\\
Total B Magnitude & 12.63 & $\nabla$ \\
Galactic Extinction $A_{B}$ & 0.519 & $\ast$\\
Angular Size & 2\farcm7 $\times$ 2\farcm3 & $\triangle$ \\
$v_{3K}$ (km s$^{-1}$) & 2428 & $\nabla$ \\
$v_{vir}$ (km s$^{-1}$) & 1972 & $\nabla$ \\
$\mu$ ($H_{0}$=65 km s$^{-1}$ Mpc$^{-1}$) & 32.41 & $\nabla$\\
\hline
\end{tabular}
\end{center}
$\otimes$ \protect\cite{will98};
$\diamond$ \protect\cite{fili98};
$\circ$ \protect\cite{woo98a};
$\odot$ \protect\cite{gall75};
$\nabla$ LEDA\footnotemark[1];
$\ast$ \protect\cite{schl98};
$\triangle$ NED\footnotemark[2]
\end{table}

\footnotetext[1]{ http://leda.univ-lyon1.fr}
\footnotetext[2]{ http://nedwww.ipac.caltech.edu}

\section{Observations and Data Reduction}

\begin{table*}
\caption{Spectroscopic observations of SN~1998A. The phase is from the
explosion epoch (JD 2450801).\label{journal}} 
\footnotesize
\begin{tabular}{cccccccc}\hline \hline
Date & JD & Phase & Telescope+Inst. & Grism or & Resol. & Exp. & Range \\
 & & (days) & & grating & (\AA) & (sec.) & (\AA) \\ \hline \hline
24/01/98 & 2450837.8 & 37 & Danish1.54+DF & grs 4,5  & 9,10 & 1800+1800 & 3380--9790 \\
26/01/98 & 2450839.8 & 39 & ESO1.52+B$\&$C & gt 15 & 9 & 2700 & 3080--10600 \\
31/01/98 & 2450844.7 & 44 & NTT+EMMI & grs 3,4 & 11,13 & 300$\times$2+300$\times$2 & 3800--10550 \\
01/02/98 & 2450845.9 & 45 & NTT+EMMI & gr 3 & 11 & 300$\times$2 & 3850--8380 \\
02/02/98 & 2450846.7 & 46 & NTT+EMMI & gr 4 & 13 & 300$\times$2 & 5680--10570 \\
18/02/98 & 2450862.7 & 62 & ESO1.52+B$\&$C & gt 15 & 11 & 3600 & 3060--10570 \\
22/03/98 & 2450894.7 & 94 & Danish1.54+DF & gr 4 & 10 & 1800 & 3420--9020 \\
27/05/98 & 2450960.5 & 160 & ESO3.6+EF2 & gr 12 & 13 & 1140 & 6070--10260 \\
29/05/98 & 2450962.6 & 162 & ESO3.6+EF2 & grs 11,12 & 13,13 & 900+900 & 3380--10290 \\
27/11/98 & 2451144.8 & 344 & ESO3.6+EF2 & gr 12 & 13 & 1800 $\times$2 & 6030--10260 \\
08/02/99 & 2451217.7 & 417 & ESO1.52+B$\&$C & gt 15 & 10 & 2700 $\times$2 & 3090-10810 \\ \hline
\end{tabular}
\end{table*}

Our photometric and spectroscopic monitoring of \98a\/ began about 2 weeks
after discovery and lasted for $\sim$ 400 days. All data
were obtained using the ESO telescopes of La Silla
(Chile), namely the Danish 1.54m + DFOSC, ESO 1.52m + B$\&$C, Dutch 0.9m, ESO
3.6m + EFOSC2, ESO NTT + EMMI, ESO 2.2m + IRAC2b.

The data reduction was performed with standard procedures within the
IRAF\footnotemark[3]
\footnotetext[3]{IRAF is distributed by the National Optical Astronomy
Observatories, which are operated by the Association of Universities
for Research in Astronomy, Inc., under cooperative agreement with the National
Science Foundation.} environment, 
including bias, overscan, and flat--fielding corrections. At early phases the
SN magnitudes were measured with a
PSF--fitting technique. However, as the SN luminosity
decreased, this method was not longer applicable. This is
because of the location of the SN, near a bright H II
region, in the southern arm of IC 2627. Therefore it was necessary to apply a
template subtraction method. This method makes use of template images of the galaxy,
which are geometrically and photometrically registered, and
seeing matched to the target image. For a detailed description
of the method, see e.g. Sollerman et al. \shortcite{soll02}.
As template images we used deep frames obtained on 2001 
February 1 (i.e. about 3 years after the explosion)
with the ESO 3.6m telescope, under excellent seeing conditions. 

The SN magnitudes were computed by reference to a local sequence of 5
stars in the field of IC 2627 (see Fig. \ref{field98A}), which in turn were calibrated
after comparison with Landolt standard stars \cite{land92}.

We note that our photometry of \98a\/ is systematically 
about 0.3 magnitudes fainter than that reported by Woodings et
al. \shortcite{woo98b} obtained with a PSF--fitting technique. 
The difference may be due to the different quality of
the images (e.g. the seeing and pixel sampling), and some subjectivity
in the measurements. We also cannot exclude intrinsic differences in the
instrumental band pass.

\newcommand{\checkit}{\fbox{{\tiny$\surd$}}}

In addition to the optical data, we observed  \98a\/ at two epochs in
the H, J, K' bands with the ESO 2.2m telescope equipped with IRAC2b. The
data were obtained when the SN was passing from the photospheric phase to
the nebular one ($\sim$ 100--120 days). The IR images were reduced
using IRAF tasks 
(including flat fielding and illumination corrections, sky
subtraction, image matching 
and stacking in order to increase the S/N ratio).
The magnitudes were estimated with a PSF--fitting method. They
were calibrated with respect to three stars in the SN
field (Fig. \ref{field98A}), observed during
the photometric night of 1998 March 28 and calibrated with infrared
standards from the IRIS photometric standard star list.
Note that the IR local stars are different from the optical ones, because
of the very small field of view of IRAC2b.  The transformation from
K' to K magnitudes was performed according to the conversion relations
given by Wainscoat $\&$ Cowie \shortcite{wain92}.

The spectra were reduced using standard IRAF tasks.  In
Tab. \ref{journal} the log of the spectroscopic observations is
shown. The spectra were wavelength calibrated by comparison with spectra
of He--Ne and He--Ar lamps and flux calibrated using spectra of standard stars
(Hamuy et al., 1992; Hamuy et al., 1994; Stone $\&$
Baldwin, 1983; Baldwin $\&$ Stone, 1984)
obtained during the same nights.
The flux calibration of the optical spectra was checked against the
photometry and, if discrepancies occurred, the spectral fluxes were
scaled to match the photometric data. The agreement with photometry was
finally within 5$\%$.

\section{Light and Colour Curves} \label{lcurv}

\begin{figure}
\includegraphics[width=9.6cm]{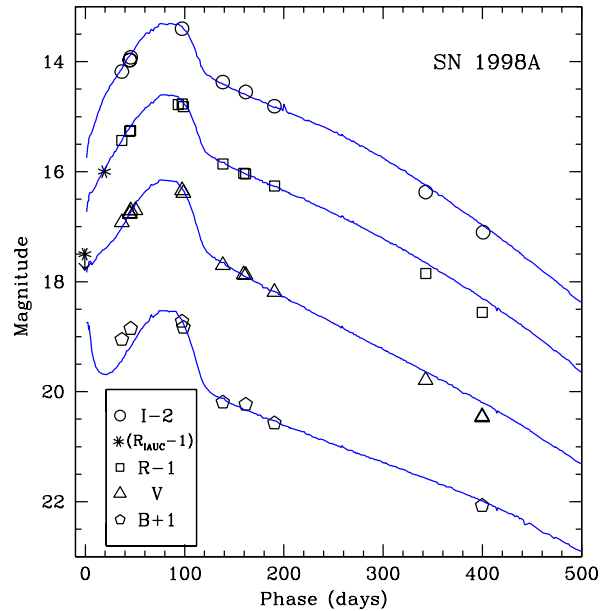}
\caption{B, V, R, I light curves of \98a. The curves of \87a
in the corresponding bands (solid lines) are also plotted, with an
offset in magnitude ($\Delta$B = 13.0, $\Delta$V = 13.2, $\Delta$R =
13.3, $\Delta$I = 13.4) to match the points of \98a\/. The asterisks
represent the prediscovery R band limit and the discovery R band photometry 
from Williams et al. (1998). The temporal fit is obtained
assuming that the explosion of SN~1998A occurred 19 days before discovery.
The adopted explosion date (phase = 0) of SN 1987A corresponds to the epoch
of the first neutrino detection JD = 2446849.82 (Bionta et al., 1987;
Hirata et al., 1987). The data of SN~1987A are from Menzies et al. (1987),
Catchpole et al. (1987), Catchpole et al. (1988), Whitelock et al. (1988),
Catchpole et al. (1989), Whitelock et al. (1989).\label{lightcurves}} 
\end{figure}

\begin{figure}
\includegraphics[width=8.1cm]{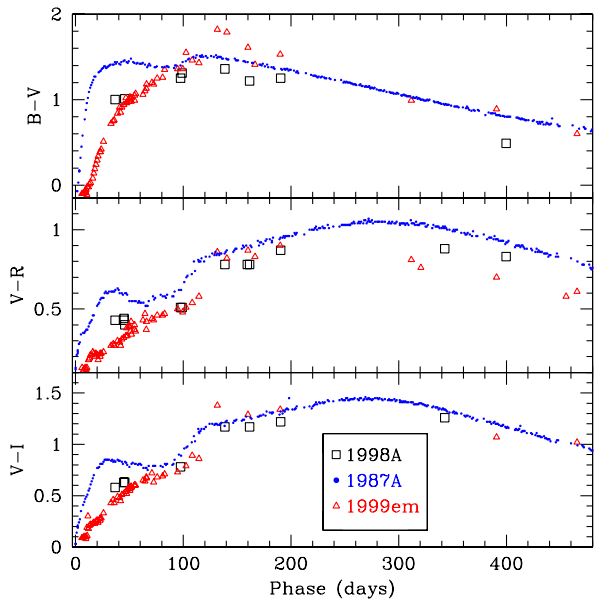}
\caption{B--V, V--R, V--I colour curves of \98a\/ and comparison with
\87a and SN~1999em. The data of SN~1999em are from Hamuy et
al. \protect\shortcite{hamu01a},
Leonard et al. \protect\shortcite{leon02}, Elmhamdi et al. \protect\shortcite{abou02}.
For SN~1999em we assume: explosion epoch JD = 2451476 and a total extinction A$_V =$ 0.31.
As in Fig. \ref{lightcurves}, the SN~1998A data are shifted in phase by +19 days from
discovery.\label{color}} 
\end{figure}

\begin{figure}
\includegraphics[width=9cm]{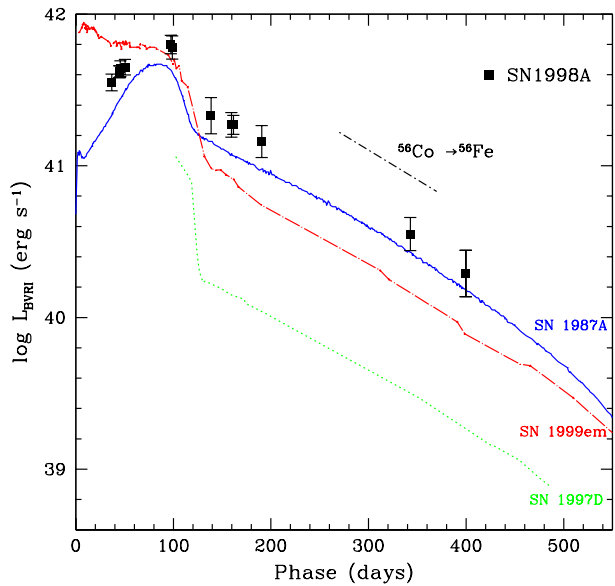}
\caption{The integrated BVRI luminosity of \98a~compared with \87a
(Patat et al., 1994, and references therein), the SN II--P 1999em
\protect\cite{abou02} and the very low--luminosity SN~1997D \protect\cite{bene01}. 
Errors  from the distance modulus estimate have not been taken into account.
\label{bolome} }
\end{figure}

\begin{table*}
\caption{Magnitudes of the sequence of local stars in the field 
of IC 2627. The errors in brackets are the r.m.s. of the 
available estimates. In the U band only the measurements of 1998 May 29 are reported. 
The IR magnitudes of 1998 March 28 are also shown.\label{std}} 
\scriptsize
\begin{tabular}{|c|c|c|c|c|c|c|c|c|} \hline \hline
Star & U & B & V & R & I & J & H & K$^\ddag$ \\ \hline \hline  
1 & 17.96 & 17.71 (0.01) & 16.96 (0.01) & 16.50 (0.01) & 16.06 (0.01) & -- & -- & -- \\
2 & 18.97 & 18.54 (0.02) & 17.69 (0.01) & 17.20 (0.01) & 16.74 (0.01) & -- & -- & -- \\
3 & 17.26 & 17.07 (0.01) & 16.35 (0.01) & 15.90 (0.01) & 15.45 (0.01) & -- & -- & -- \\
4 & 17.79 & 17.61 (0.01) & 16.88 (0.01) & 16.41 (0.01) & 15.89 (0.01) & -- & -- & -- \\
5 & 16.66 & 16.65 (0.01) & 16.06 (0.01) & 15.67 (0.01) & 15.25 (0.01) & -- & -- & -- \\
A & -- & -- & -- & -- & -- & 15.19 & 14.54 & 14.67 \\
B & -- & -- & -- & -- & -- & 16.28 & 15.52 & 15.63 \\
C & -- & -- & -- & -- & -- & 17.08 & 15.96 & 15.96 \\ \hline
\end{tabular}

$^\ddag$ The K magnitudes are obtained from K'
measurements using the conversion relations by Wainscoat $\&$ Cowie
(1992)
\end{table*}

\begin{table*}
\caption{Optical and IR photometry of \98a\/ (JD +2400000).} \label{photometry}
\scriptsize
\begin{tabular}{|c|c|c|c|c|c|c|c|c|c|c|} \hline \hline
Date & JD & Phase & B & V & R & I & J & H & K & Source \\ \hline \hline 
18/12/97 & 50800.5 & -0.5 & -- & -- & $\geq$18.5 & -- &  -- & -- & -- & 0 \\
06/01/98 & 50820.3 & 19 & -- & -- & 17.0 & -- &  -- & -- & -- & 0 \\
24/01/98 & 50837.8 & 37 & 18.05 (0.10) & 16.92 (0.03) & 16.43 (0.04) & 16.18 (0.03) & -- & -- & -- & 1 \\
01/02/98 & 50845.9 & 45 & -- & 16.76 (0.02) & 16.27 (0.01) & 15.97 (0.01) & -- & -- & -- & 5 \\
02/02/98 & 50846.7 & 46 & 17.85 (0.08) & 16.71 (0.04) & 16.25 (0.01) & 15.92 (0.06) & -- & -- & -- &  2 \\
22/03/98 & 50894.7 & 94 & -- & -- & 15.78 (0.01) & -- &  -- & -- & -- & 1 \\
26/03/98 & 50898.5  & 98 & 17.72 (0.11) & 16.34 (0.05) & 15.77 (0.04) & 15.40 (0.05) & -- & -- & -- &  2 \\
27/03/98 & 50899.8 & 99 & 17.83 (0.17) & 16.39 (0.03) & 15.82 (0.06) & -- & -- & -- & -- &2 \\
28/03/98 & 50900.8 & 100 & -- & -- & -- & -- & 15.17 (0.03) & 14.66 (0.04) & 14.52 (0.05) & 4 \\ 
15/04/98 & 50918.8 & 118 & -- & -- & -- & -- & 15.66 (0.07) & 15.04 (0.06) & -- & 4 \\
27/04/98 & 50930.6 & 130 & -- & -- & -- & -- & 15.40 (0.06) & 14.94 (0.09) & 14.94 (0.16) & 6 \\   
06/05/98 & 50939.6 & 139 & 19.19 (0.28) & 17.70 (0.11) & 16.86 (0.08) & 16.37 (0.08) &  -- & -- & -- & 2 \\
27/05/98 & 50960.5 & 160 & -- & 17.87 (0.10) & 17.03 (0.06) & -- & -- & -- & -- & 3 \\
29/05/98$^{(\ast)}$ & 50962.5  & 162 & 19.23 (0.10) & 17.88 (0.05) & 17.04 (0.05) & 16.55 (0.05) &  -- & -- & -- & 3 \\  
05/07/98 & 50991.5  & 191 & 19.57 (0.16) & 18.19 (0.09) & 17.26 (0.02) & 16.81 (0.14) & -- & -- & -- &  2 \\
26/11/98 & 51143.8 & 343 &  -- & 19.79(0.02) & 18.85 (0.03) & 18.37 (0.07) &  -- & -- & -- & 3 \\
21/01/99 & 51200.7 & 400 & 21.07 (0.32) & 20.45 (0.02) & 19.56 (0.03) & -- &  -- & -- & -- & 1 \\
22/01/99 & 51201.7 & 401 & -- & -- & -- & 19.10 (0.18) &  -- & -- & -- & 1 \\
01/02/01 & 51941.8 & 1141 & -- & $\geq$23.2 & $\geq$22.9 & -- &  -- & -- & -- & 3 \\ \hline
\end{tabular}

0 = IAU Circ. 6805; 1 = Danish 1.54m + DFOSC; 2 = Dutch 0.90m; \\ 
3 = ESO 3.6m + EFOSC2; 4 = ESO 2.2m + IRAC2b; 5 = ESO NTT + EMMI; 6 = IAU Circ. 7435 \\
$^{(\ast)}$ U = 19.29 $\pm$ 0.30
\end{table*}
 
The photometric monitoring of \98a\ covers a period of more than one year
and, although not densely sampled, it is sufficient to describe the
overall evolution of the object. The magnitudes of the local standards
are reported in Tab. \ref{std}, while the optical and IR photometry is
reported in Tab. \ref{photometry}. The B, V, R, I light curves are
shown in Fig. \ref{lightcurves}.  The photometric evolution of SN~1998A
is very similar to that of \87a.  Using the light curve of SN 1987A
\cite{menz87,catc87,catc88,whit88,catc89,whit89} as a template, we
adjusted the explosion epoch of SN~1998A to get the best match between
the light curves of the two objects. Consistently with Woodings et
al. \shortcite{woo98b}, 
we deduce that \98a~exploded
$\sim$ 19 days before discovery, close to the epoch of the
prediscovery limit of Williams et al. \shortcite{will98}. Hereafter
the phases will be relative to this epoch (JD = 2450801).  We expect
that, similar to SN~1987A, a decrement of the B band luminosity after
shock breakout has  occurred, but unfortunately we have no direct
evidence for it.  The initial UV excess and the subsequent observed dip
in U and B bands observed in SN 1987A (e.g. Hamuy et al., 1988), in SN
1993J (e.g. Schmidt et al., 1993), and recently in the Type Ib/c SN
1999ex \cite{stri02} are probably common features in core--collapse
(CC) SNe.

Later on the SN luminosity begins to rise reaching a broad maximum
during which, in analogy to SN~1987A, we expect that the recombination of the
hydrogen envelope takes place. This phase corresponds to the plateau in 
``normal'' SNe II--P.  

After that, the luminosity drops
abruptly, until the main contribution to the flux comes from the
radioactive decay $^{56}$Co $\rightarrow$ $^{56}$Fe: the nebular
phase begins.

We note that at t $\sim$ 300 days past explosion, the light curve
becomes steeper than the early radioactive tail, indicating an
increasing escape of trapped $\gamma$--rays and/or dust formation.
The measurements of the slopes of the early radioactive tail (between
$\sim$ 130 and 200 days) give values: $\gamma_B \approx$ 0.75,
$\gamma_V \approx$ 0.95, $\gamma_R \approx$ 0.77, $\gamma_I \approx$
0.85~mag/100$^{d}$ which are slightly smaller than the slope of the
radioactive decay of $^{56}$Co (0.98 mag/100$^{d}$), while between 330
and 410 days they are larger than expected from the $^{56}$Co decay:
$\gamma_V' \approx$ 1.18, $\gamma_R' \approx$ 1.25 and $\gamma_I'
\approx$ 1.26~mag/100$^d$. We note that the steepening in the late
time light curve occurs earlier in \98a\/ than in \87a\/.

The overall behaviour of the colour curves of \98a\/
(Fig. \ref{color}) is rather similar to \87a, although \98a\/~is bluer,
particularly at early epochs. This indicates
higher temperatures for \98a\/ at comparable phases, as also confirmed by the spectra (see
Sect. 4). The colour evolution of SN~1998A is also similar to that of the normal
plateau SN~1999em \cite{hamu01a,leon02,abou02} (Fig. \ref{color}).

In order to compute the absolute luminosity, we assume a distance
modulus $\mu$ = 32.41 for IC 2627 (obtained via recession velocity corrected by
Local Group infall into the Virgo cluster, with H$_{0}$ = 65 km s$^{-1}$
Mpc$^{-1}$; see also Tab. \ref{hostgal}).  We also adopt the value of
foreground Galactic extinction at the coordinates of the SN from 
Schlegel et al. (1998).  No interstellar absorption
lines are seen in the spectra of \98a\/ at the host galaxy rest wavelength. 
This was also found by Leonard $\&$ Filippenko
\shortcite{leon01} who evaluated an upper limit to the equivalent
width of interstellar Na ID absorption and concluded that the
host--galaxy reddening was small (E$_{B-V} \le$ 0.03). 
Therefore, we also assume the host galaxy extinction to be negligible
and adopt an extinction entirely due to the Galaxy (A$_{B,TOT}$ = 0.52).

Guided by the resemblance of the light curve to SN~1987A
(Fig. \ref{lightcurves}), we can estimate the absolute 
luminosity of SN~1998A at the broad peak as 
M$_B\sim -$15.4, M$_V\sim -$16.6, M$_R \sim -$17.1, M$_I \sim -$17.4.

In Fig.~\ref{bolome}, the pseudo--bolometric BVRI light curve of \98a~is
shown compared with those of SN~1987A, SN 1999em [data from Elmhamdi 
et al. \shortcite{abou02}, with distance modulus obtained from
Cepheids calibration by Leonard et al. \shortcite{leon03}]
and SN 1997D \cite{bene01}.
We note that at late times \98a~is brighter than \87a\/.
From a comparison between the luminosities of
\98a~and \87a during the early radioactive tail phase, we estimate that 0.11
M$_{\odot}$ of $^{56}$Ni have been ejected by the explosion (M$_{Ni}$ = 0.075 M$_{\odot}$ 
in \87a\/). 

\98a\/ was mostly monitored in the BVRI bands, but a few sparse
U--band and near--IR observations allow us to compute the contribution
of these bands to the total flux at the beginning of the nebular
phase.  We estimate that the U--band flux on 1998 May 29 (phase
$\sim$ 162 days) contributes only to $\approx$ 7$\%$ of the total
optical ({\sl UBVRI}) flux, in accordance with the red colour of the
SN.  We used the near infrared magnitudes to estimate the IR
contribution after recombination of the H envelope. The JHK flux is
comparable to the optical (about 75$\%$ of the BVRI flux).  At
comparable phases, \87a showed an IR flux (extended also to the L
band) $\approx$ 130$\%$ of the optical one \cite{catc87}.  Therefore,
at least on these epochs, the contribution of IR bands to the total
flux is significant.
\section{Spectroscopy}

\subsection{Spectroscopic Evolution and a Comparison with SN 1987A}
\subsubsection{Photospheric Spectra}
\begin{figure*}
\includegraphics[width=15.2cm,angle=0]{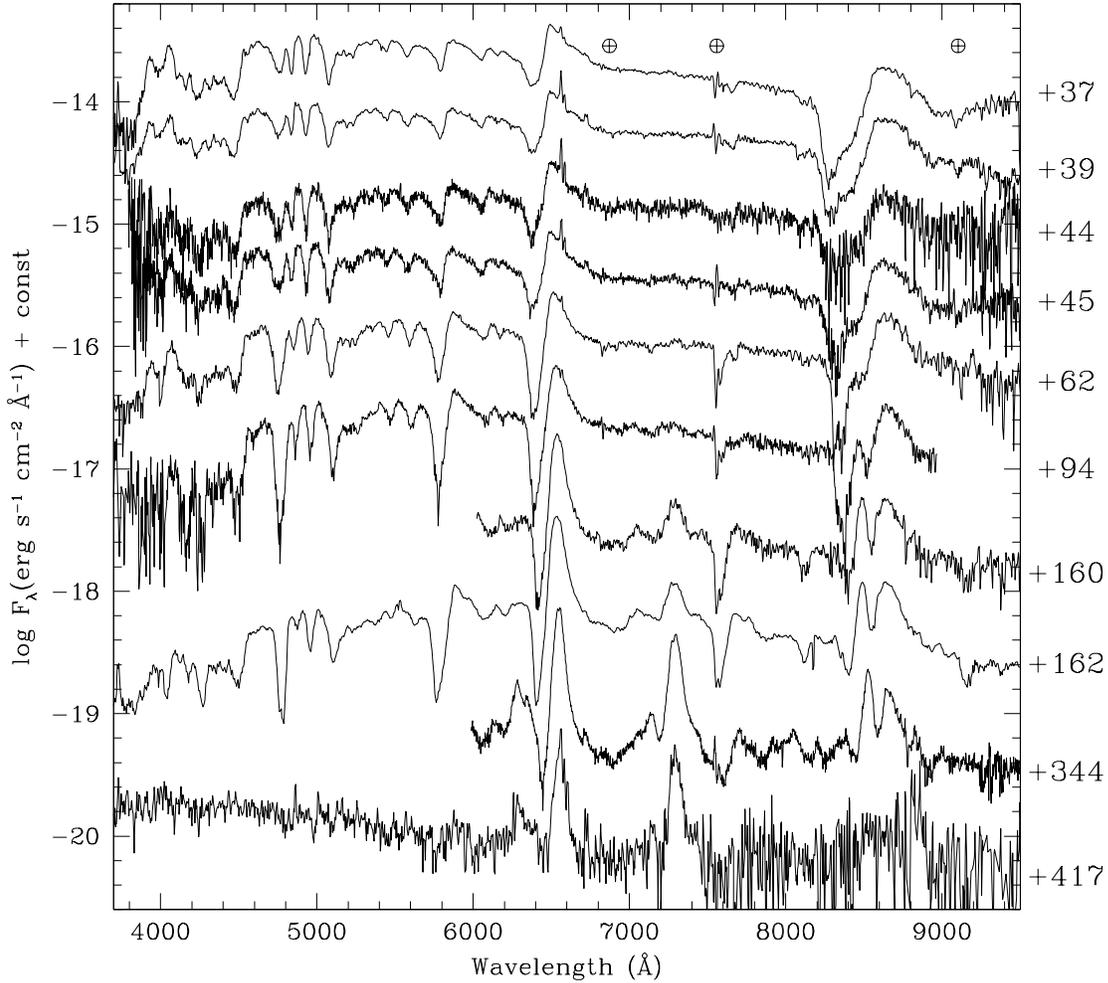}
\caption{The spectroscopic evolution of \98a~from the photospheric to
the nebular phase.
The labels to the right indicate the days after the explosion. 
The spectrum at +45 days results from the combination of the spectra obtained on 1998 February
1 and 2.
The contamination of an H II region is visible from the presence of narrow H$\alpha$, [N II], [S II]
lines in the region 6550--6700 \AA. The position of telluric bands,
partially removed, is labelled by $\oplus$.
All spectra are redshift corrected.} \label{evoluzione98A}
\end{figure*}

Our spectroscopic data of \98a~cover a period of $\sim$ 400 days.  The
entire spectroscopic evolution is shown in Fig. \ref{evoluzione98A}.
The first spectrum was obtained about 18 days after discovery (phase $\sim$ 37 days).
It shows the photospheric broad P--Cygni lines of H I, Ca II, Na ID, Fe II, Ti II
and many other metal lines typical of CC--SNe during the recombination
phase. 
A more extensive line identification will be given in Sect. \ref{lineid98A}.

\begin{figure*}
\includegraphics[width=16.5cm]{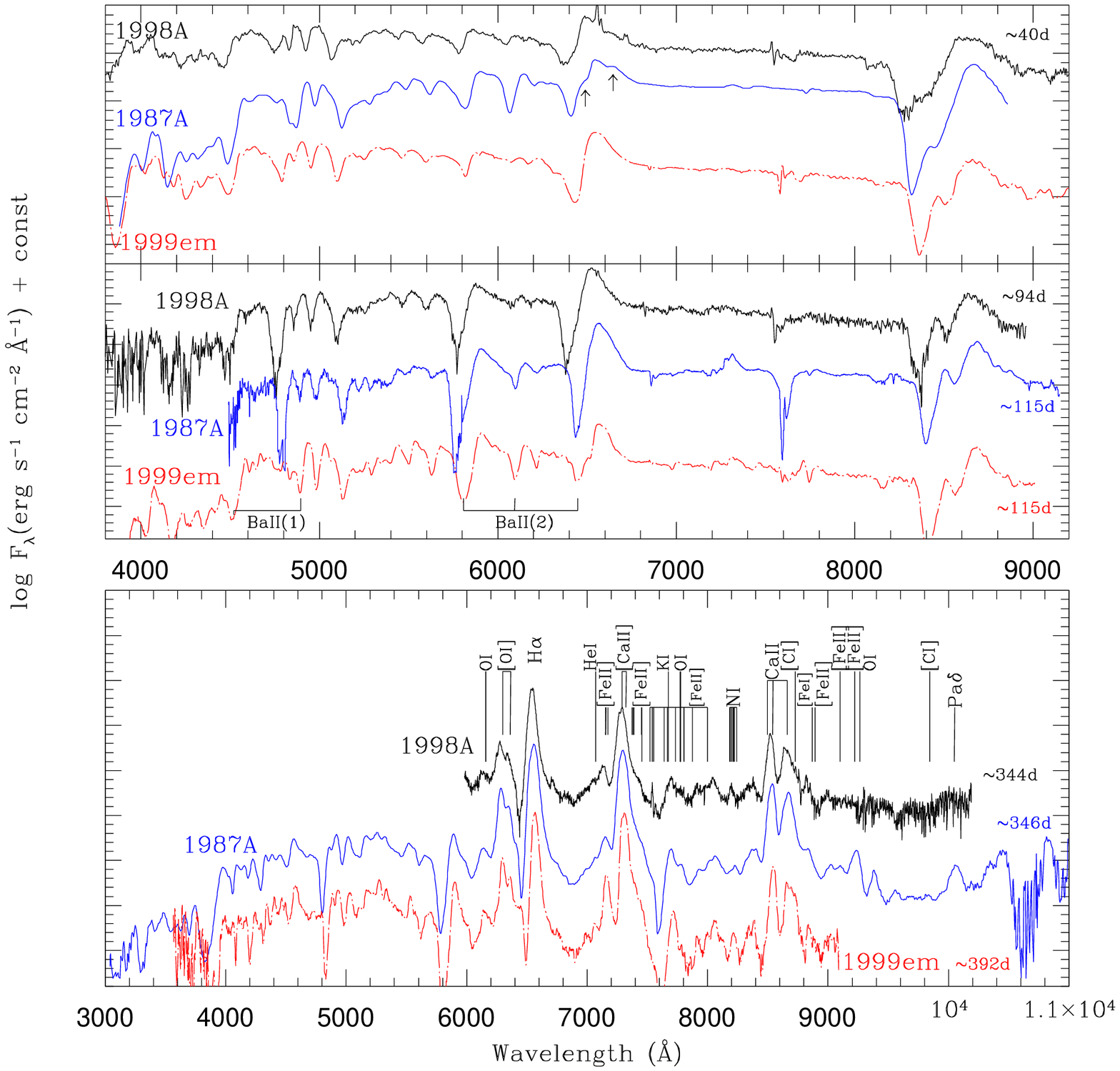}
\caption{Spectra of \98a\/, \87a and SN~1999em at comparable phases. A strong
resemblance among these spectra is evident. In the spectrum of
SN~1987A at phase $\sim$40 days, the position of the bumps in the H$\alpha$ 
profile ({\sl Bochum event}) is labeled by arrows.
In the late--photospheric spectrum of SN~1999em the position
of the minima of main Ba II lines is marked. 
In the nebular spectrum of \98a\/ (taken about 1
year after the explosion) the main emission features are
marked. The spectra of SN~1999em are from Elmhamdi et al. (2003), those
of SN~1987A from the Padova--Asiago SN Archive.
\label{cfr_87A}} 
\end{figure*}

As already noted by Leonard $\&$ Filippenko \shortcite{leon01}, the 
spectroscopic evolution of \98a\/ (in analogy with \87a) is surprisingly rapid.
On day $\sim$ 30 the spectrum already showed strong metal lines. 
A comparison with a spectrum of SN 1999em at
phase $\sim$ 35 days shows that \98a\/ exhibited stronger metal features \cite{leon01}.
In Fig. \ref{cfr_87A} we compare spectra of SNe~1998A, 1987A and 1999em at similar
ages. It is remarkable at $\sim$ 40 days post--explosion (just 5 days after the 
comparison reported by Leonard $\&$ Filippenko) the overall similarity among the
spectra of the three SNe. 
 
In analogy with \87a, also \98a\/ probably experienced rapid colour
evolution and a strong deficit of flux at
short wavelengths, due to the rapid adiabatic cooling of the
envelope and, at $\lambda \le$ 4500 \AA, to line blanketing.  

However, from Fig. \ref{cfr_87A} we note also a number of interesting
differences between SNe~1998A and 1987A.  The Ba II lines of SN~1998A
are similar in strength to those of SN 1999em at 40 days and much
weaker than those of SN~1987A.  Note also that the {\sl Bochum event},
i.e. the fine structure developed by the H$\alpha$ profile and
detected in the spectra of \87a in a phase comparable to that of our
first spectra of \98a\/~\cite{hanu87}, is not visible in \98a\/ at any
epoch (see Fig. \ref{cfr_87A} -- Top).

\begin{figure}
\includegraphics[width=9cm]{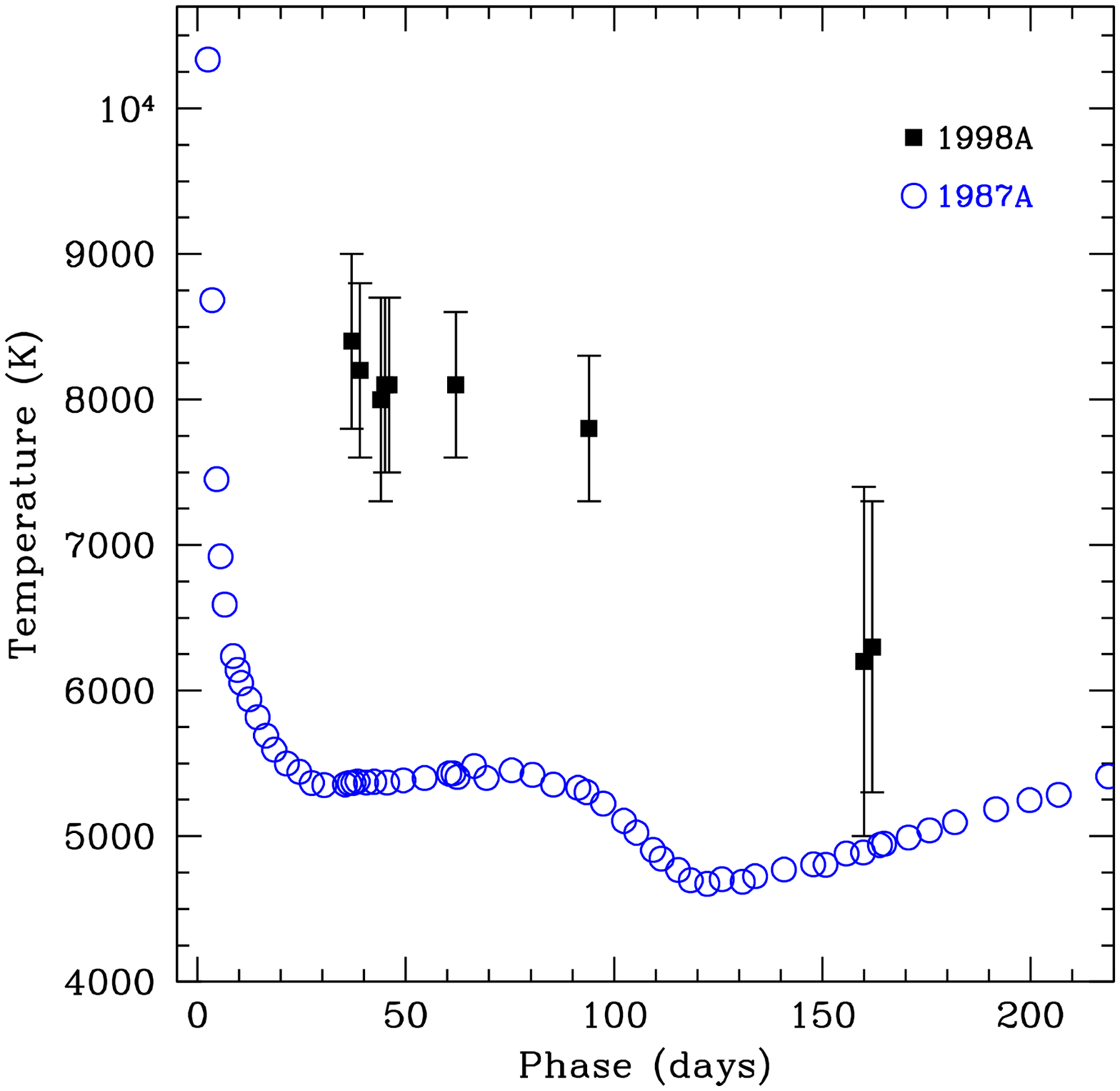}
\caption{Evolution of the temperature from the blackbody fit of the
continuum and comparison with \87a. The data of \87a are from Phillips
et al.~(1988). \label{temperature}}
\end{figure}

The temperature deduced by a blackbody fit of the continuum
(see Fig. \ref{temperature}) is about 8200 K at $\sim$ 40 days. This
is perhaps higher 
than expected given the presence of a number of metal 
lines and definitely higher than that of \87a at the same phase.
As expected, the temperature evolution measured on the plateau is
slow.

\begin{figure}
\includegraphics[width=9cm]{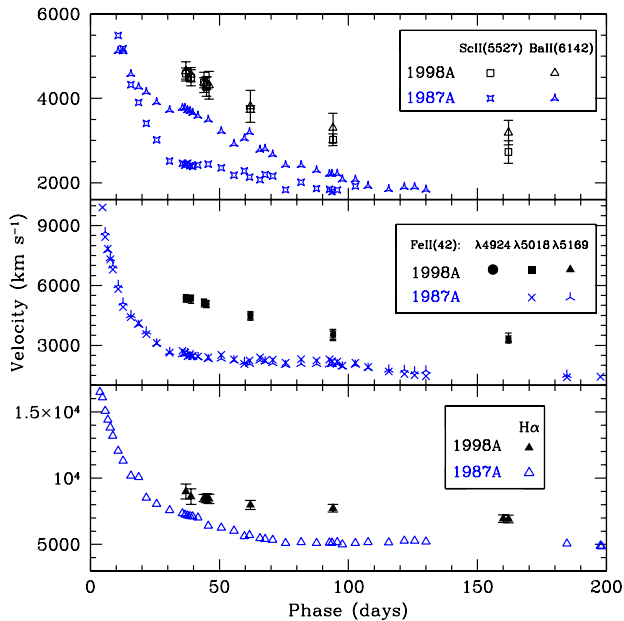}
\caption{Comparison between the expansion velocities of \98a\/ and \87a. 
The velocities are derived from the minima of Ba II $\lambda$6142 and Sc II
$\lambda$5527 ({\bf Top}); Fe II multiplet 42
$\lambda\lambda\lambda$ 4924, 5018, 5169 
({\bf Centre}); H$\alpha$ ({\bf Bottom}). Again the 
data of \87a are from Phillips et al.~(1988).\label{velocity98A}} 
\end{figure}

The photospheric velocity estimated by the minima
of Sc II 5527 \AA~ and/or Ba II 6142 \AA~ 
varies from about 4500 \kms~at $\sim$ 40 days to
3000--3200 \kms~at phase $\sim$ 94 days (see Fig. \ref{velocity98A} -- Top),
considerably higher than that of \87a at comparable phase.
The velocity of the Fe II lines (multiplet 42 lines, Fig. \ref{velocity98A} -- Centre)
shows the same behaviour, as well as the velocity derived from H$\alpha$ 
(Fig. \ref{velocity98A} -- Bottom).

\begin{figure}
\includegraphics[width=8.5cm]{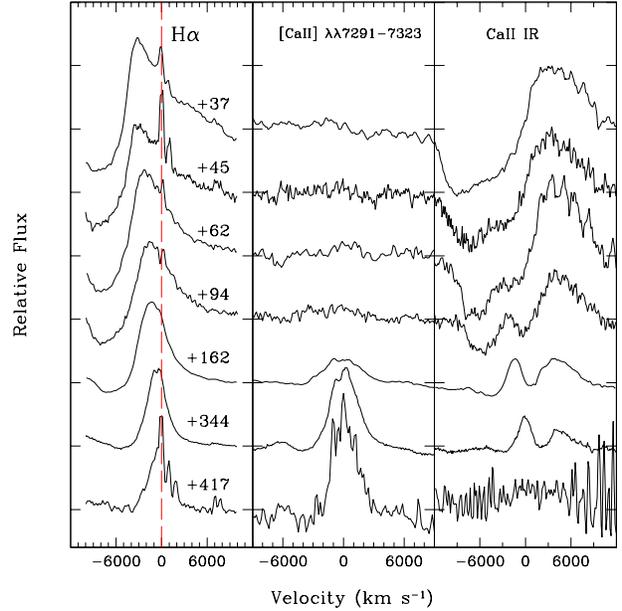}
\caption{Evolution of the profile of some spectral 
lines of SN~1998A. All spectra have been normalised to the peak 
of the broad H$\alpha$. \label{mainlines}}
\end{figure}

\begin{figure}
\includegraphics[width=8.7cm]{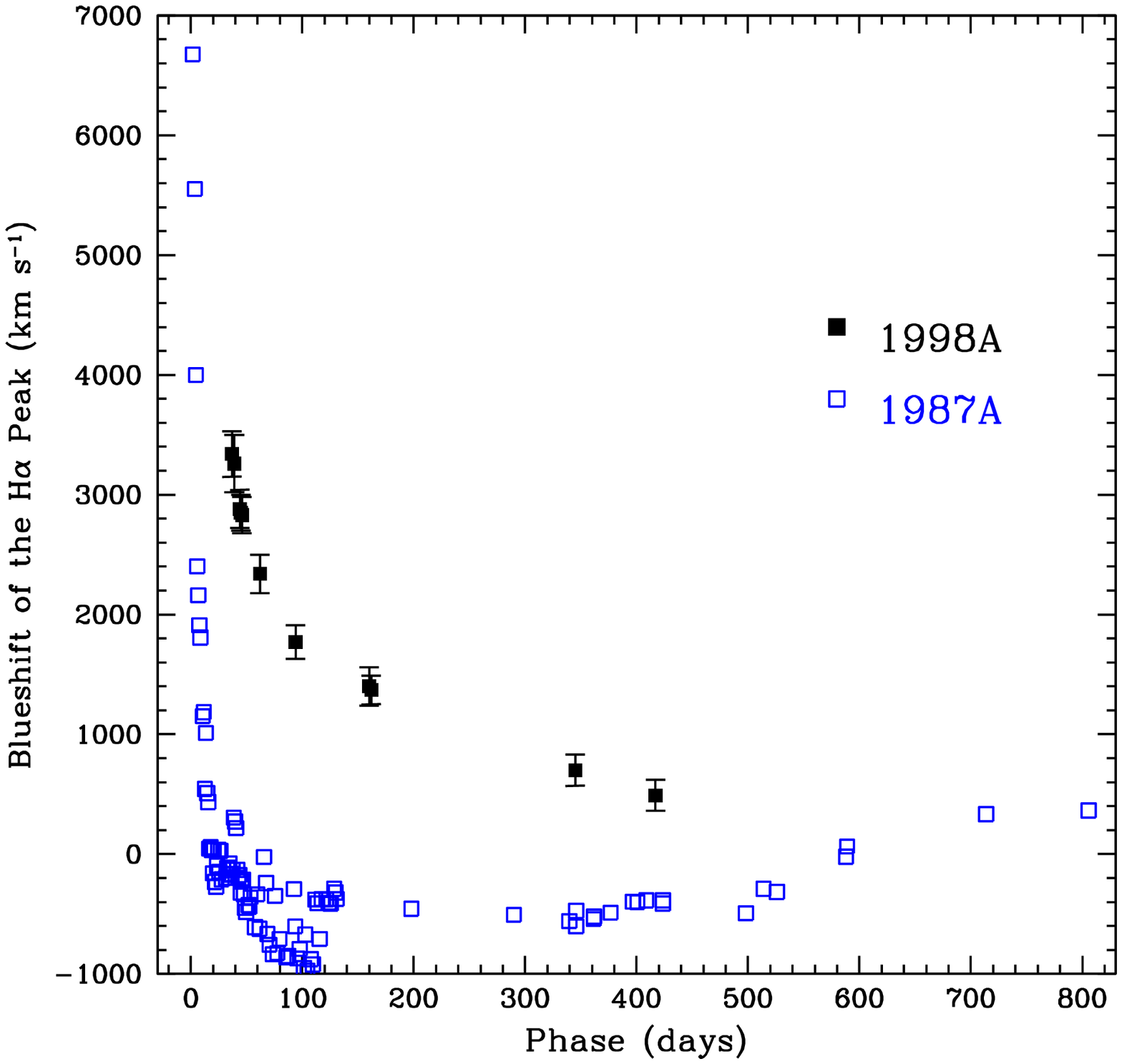}
\caption{Comparison between \98a~ (filled symbols) and \87a (empty
symbols): blueshift of the H$\alpha$ emission peak
with respect to the rest wavelength. The data for \87a are taken from
Phillips et al.~(1988) and Phillips et al.~(1990).\label{halpha}} 
\end{figure}

\subsubsection{Nebular Spectra}
The spectrum at phase $\sim$ 94 days shows the emergence of [Ca II]
7291--7324 \AA\/ nebular lines, that become systematically stronger in
the latest spectra.  In the spectrum at phase 160 days, at the same
epoch of the steep luminosity drop that marks the beginning of the
nebular phase, we note evidence of a decrease in temperature (T $\sim$
6200 K).

The spectrum at day~345 shows the typical nebular features of a SN II:
Balmer (and possibly Paschen) lines of H, Ca II IR, [Ca II] 7291--7324
\AA, [O I] 6300--6364 \AA\/ doublet, and several line blends of [Fe
II], the strongest of which is the feature near 7200 \AA~(see e.g. 
Spyromilio et al., 1991).  We
tentatively identify the line at about 6150 \AA\/ as O I and the
feature at 8200 \AA\/ as N I. At redder wavelengths the Paschen lines
may contribute to the strength of some features: in particular, the
peak at $\lambda$ 10010 \AA\/ might be Paschen $\delta$. More doubtful
is the presence of [C I] lines, possibly blended with Ca II IR and [Fe
II].

Our final, low S/N spectrum, obtained about 420 days after the
explosion, shows intense nebular features of [Ca II] and [O I]. The Ca
II IR has become weaker than the [Ca II] 7291--7324 \AA\/
doublet. The observed blue pseudo--continuum is not intrinsic but
probably due to background contamination by hot, young stars
associated with H II regions close in projection to \98a\/.

\subsubsection{Blueshift of H$\alpha$ emission and Ca II IR triplet evolution}

Fig.~\ref{mainlines} shows the evolution of H$\alpha$, [Ca II] and Ca
II IR in the spectra of \98a\/ (corrected to the host galaxy rest
frame).  The narrow H$\alpha$ line emitted by the underlying H II
region visible in several spectra shows the significant
blueshift of the SN H$\alpha$ emission at early--times.  This effect
was seen also in \87a and in other SNe II, e.g.  SN~1988A
\cite{tura93} and SN~1999em \cite{abou02}.  As shown in
Fig. \ref{halpha}, the blueshift of H$\alpha$ in \98a\/~is
significantly larger than that observed in SN~1987A.  In \87a a
blueshift of the H$\alpha$ emission component of about 6000 km
s$^{-1}$ was detected soon after the explosion (Menzies et al., 1987),
but it decreased rapidly, disappearing about 20 days after the
explosion.  Chugai \shortcite{chug88} explained this initial blueshift
of the H$\alpha$ peak as due to diffuse reflection of resonance
photons emitted inward toward the photosphere.  In \98a\/ the
blueshift is detectable until $\sim$ 1 year after the explosion and
thus is more difficult to explain with this mechanism.  Other very
late spectra of CC--SNe show evidence of blueshifted emission peaks
(e.g. SN~1993J, Matheson et al., 2000).
 
In fact, \87a showed an  increase of the 
blueshift of H$\alpha$ and other forbidden lines at
very late--time (phase $\ge$ 580 days). This occurred in coincidence 
with the deviation of the light curves from the radioactive decay of $^{56}$Co 
that possibly indicates dust formation \cite{koza91}. No clear evidence of dust formation 
is visible in the late optical spectra of \98a\/. In particular there is no evidence 
of blueshift of the [O I] 6300-6364 \AA\/ emission (e.g. SN~1999em,
Elmhamdi et al., 2002). 

Also interesting is the evolution of the Ca II IR triplet.  In the
early spectra only the broad Ca II $\lambda$8662 line is detected. At
the end of recombination  the line at 8542 \AA\/ appears, and then
becomes prominent in the nebular spectra. The emission line at 8542
\AA\/ is narrower ($FWHM \sim$ 1700 \kms) than that at 8662 \AA\/
($FWHM \sim$ 2750 \kms) and the [Ca II] doublet
$\lambda$$\lambda$7291--7324 ($FWHM \sim$ 3100 \kms).  Swartz et
al. (1989) suggest that the relatively narrow width of the 8542 \AA\/
line observed in the nebular spectra of SN 1987A may have been caused
by a partial absorption by the line at 8662 \AA. Indeed, the first
line of the triplet ($\lambda$8498) is not visible at any epoch in the
spectra of SN~1998A, due to the absorption of the two redder lines of
the Ca II IR triplet \cite{li93}.  This may also explain the absence
of the O I $\lambda$8446 line.

In the next Section we will show that the broad profile of the whole Ca II IR triplet
can be reproduced without invoking blends with other lines, 
but simply with a larger expansion velocity of Ca--rich ejecta (see model on Sect.
\ref{lineid98A} and Fig. \ref{synow1}, Left--Bottom panel).

\subsection{Line Identifications: a {\sl SYNOW} Synthetic Photospheric Spectrum} \label{lineid98A}

\begin{figure*}
\includegraphics[width=10.6cm,angle=270]{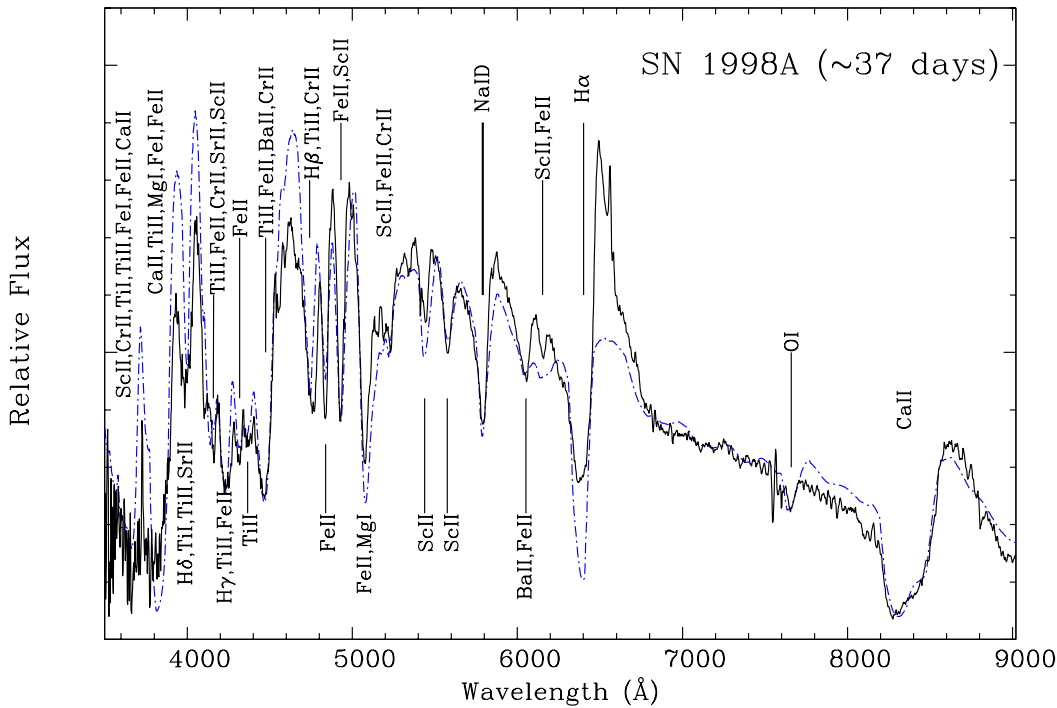}
\caption{Comparison of the observed spectrum of \98a\/ at day 37 (solid line) with the
{\sl SYNOW} synthetic spectrum (dash--dotted line). The synthetic spectrum was obtained assuming
T$_{bb}$ = 8400 K, v$_{ph}$ = 5400 km s$^{-1}$ and adjusting the opacity from 13
different ions.\label{synow}} 
\end{figure*}

\begin{figure*}
\includegraphics[width=10.7cm,angle=270]{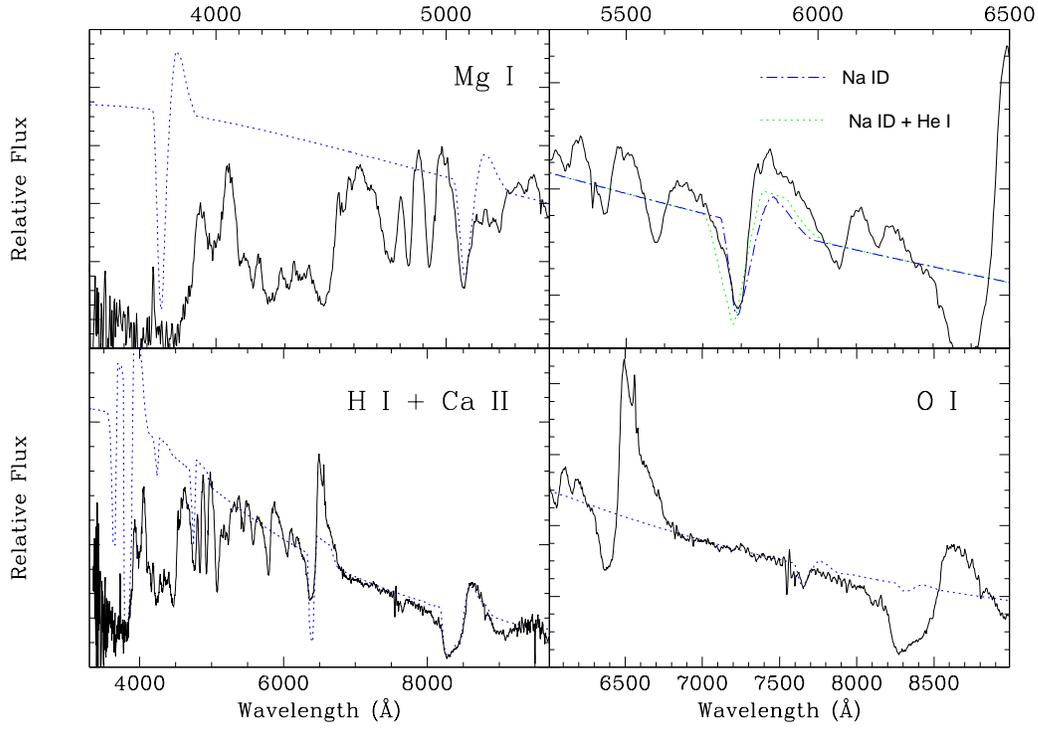}
\caption{Same as Fig. \ref{synow}, but considering only contribution of
Mg I, Na ID (\& NaID + He I), H I + Ca II, O I in the {\sl SYNOW} synthetic spectrum. 
\label{synow1}} 
\end{figure*}

\begin{figure*}
\includegraphics[width=10.7cm,angle=270]{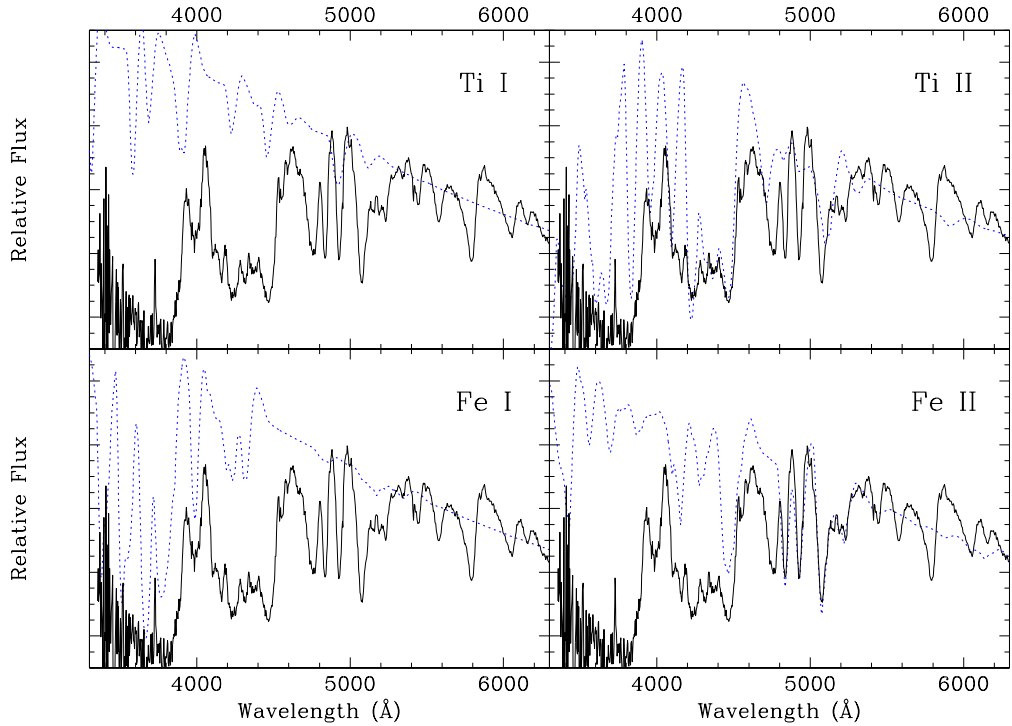}
\caption{Same as Fig. \ref{synow}, but considering only contribution of
Ti I, Ti II, Fe I, Fe II in the {\sl SYNOW} synthetic spectrum. \label{synow2}}
\end{figure*}

\begin{figure*}
\includegraphics[width=10.7cm,angle=270]{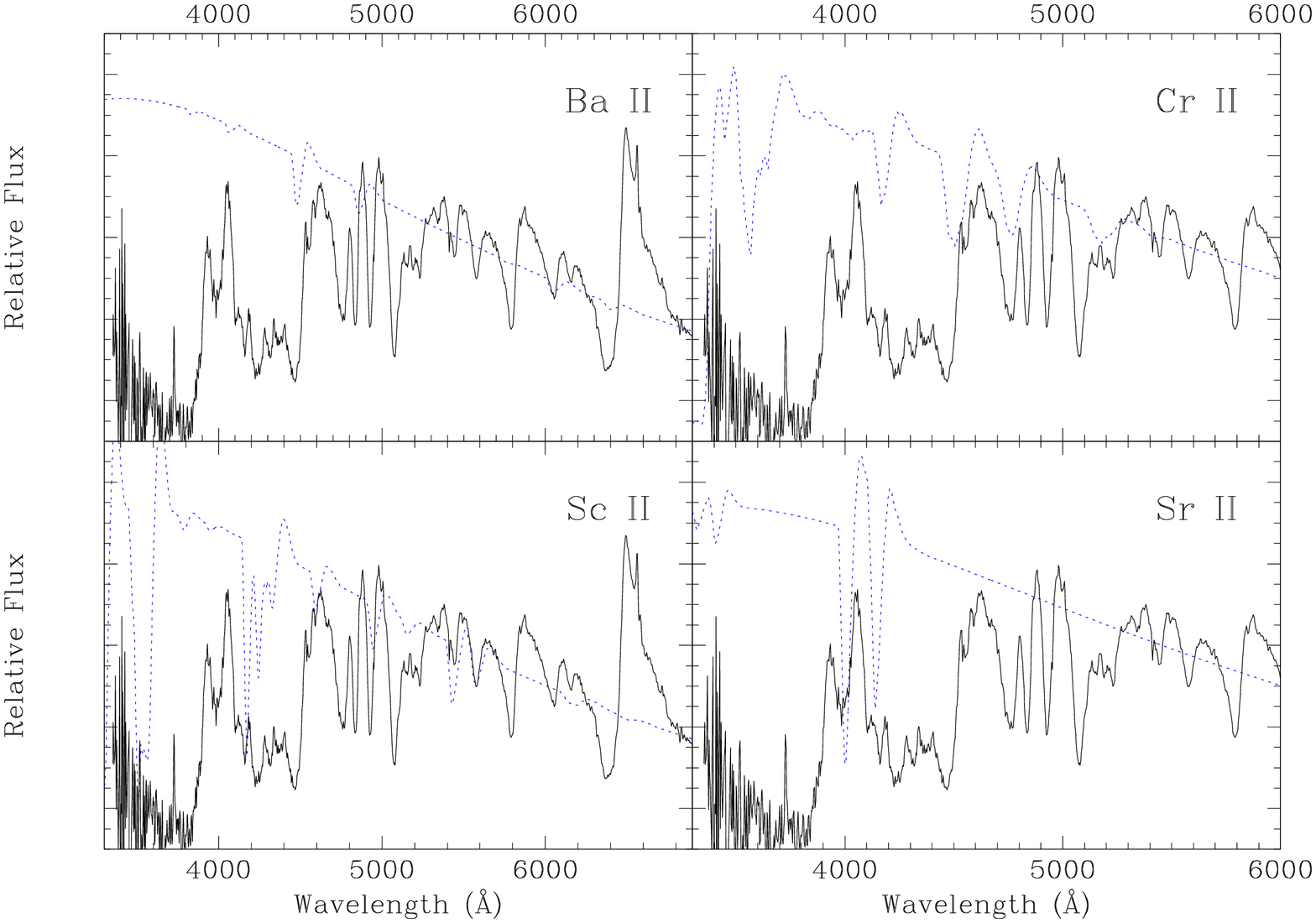}
\caption{Like Fig. \ref{synow}, but considering only contributions of
Ba II, Cr II, Sc II, Sr II in the {\sl SYNOW} synthetic spectrum.  
\label{synow3}}
\end{figure*}

In order to attempt a complete line identification, we have used the simple, parameterized code
{\sl SYNOW} \cite{fish00} to calculate a synthetic photospheric spectrum to be compared with
our observed spectrum at phase $\sim$ 37 days.  Because of the rapid spectroscopic 
evolution of \98a\/, the observed spectrum already shows the main 
P--Cygni permitted lines of metals.
The simple assumptions on which the code is based are: spherical
symmetry, homologous expansion, sharp
photosphere and line formation coming from resonant scattering.
Optical depths for the lines are calculated in the Sobolev approximation, and
free parameters are: the continuum blackbody temperature (T$_{bb}$),
the excitation temperature (T$_{exc}$) of each ion, the velocity at
the photosphere (v$_{ph}$), and the optical depths ($\tau$) of
reference lines for each ion \cite{hata99}. The relative line
intensities of an ion are adjusted by assuming LTE \cite{jeff90}.  To
fit the absorption line profiles, we assume that the radial dependence
of the optical depths goes as a power--law function of index n = 6.  In
Fig. \ref{synow} the observed spectrum at day 37 (corrected for a
redshift z = 0.0072 and for a reddening E$_{B-V}$ = 0.12) is compared with
the synthetic spectrum obtained with parameters T$_{bb}$ = 8400 K,
v$_{ph}$ = 5400 km s$^{-1}$. The ions included in the synthetic spectrum
are H I, Ca II, Na I, Fe II, Ti II, Sc II, Ba II, Sr II, Cr II, Mg I,
Fe I, Ti I, O I.  Only the lines of H I and Ca II are detached from the photosphere,
respectively at 7500 and 6300 km s$^{-1}$. We assume that T$_{exc}$
for each ion is close to 5000 K for neutral species
and 10000 K for once ionized atoms as suggested by Hatano et
al. (1999).  We adopt a T$_{bb}$ that is unusually high for a
spectrum showing such well developed permitted metal lines. Moreover,
in order to improve the quality of the fit, we also assume large optical
depths (log $\tau$ $\approx$ 2.5) for the Ti II lines.

The fit of the observed spectrum on day 37 is rather good
(Fig. \ref{synow}), with the usual exception of the H$\alpha$ emission
for which collisional excitation and other non--LTE effects are
expected to be important. Detail of the contribution of each species
is shown in Figs.~\ref{synow1}--\ref{synow3}.  The fit of Na ID is
relatively poor; we tried to improve the result using a small
contribution of He I $\lambda$ 5876, and the comparison between the
two different choices is shown in Fig. \ref{synow1}. Features
due to He I are expected to be important during the
photospheric phase only shortly after the explosion. However, the high
value of T$_{bb}$ may indicate high $\gamma$--ray deposition which
could enhance the He I $\lambda 5876$ line.  In \87a the He I line was
visible only for 3 days after the explosion and a few days later Na
ID began to emerge (see e.g. the sequence of early spectra in
Hanuschilk $\&$ Dachs, 1988; Phillips et al., 1988; Tyson $\&$
Boeshaar, 1987; Menzies et al., 1987), although in \98a\/, because of
the higher temperatures (or higher $\gamma$--ray deposition), the
situation could be different.

The line--blanketing below $\sim$ 5100 \AA\/ mainly derives from the 
strength of a number of Fe II and Ti II multiplets. To
improve the quality of the fit in the bluest region of the spectrum,
strong contributions from Ca II H $\&$ K, Sc II, Sr II and Cr II are
needed.  Also the presence of neutral species (e.g. Ti I
and Fe I) at the photospheric velocity,  improves the
quality of the fit in the line blanketed region.  We stress that
lines of neutral ions are expected to be produced in
external regions (if at all),  with velocities larger than v$_{ph}$. 
The absorption feature at $\lambda \sim$ 7700 \AA\/ is
reproduced by O I and the P--Cygni profile of
the strong Ca II IR triplet is well fitted by the synthetic spectrum.

\subsection{A \texttt{PHOENIX} Model}

\begin{figure*}
\includegraphics[width=9.6cm,angle=90]{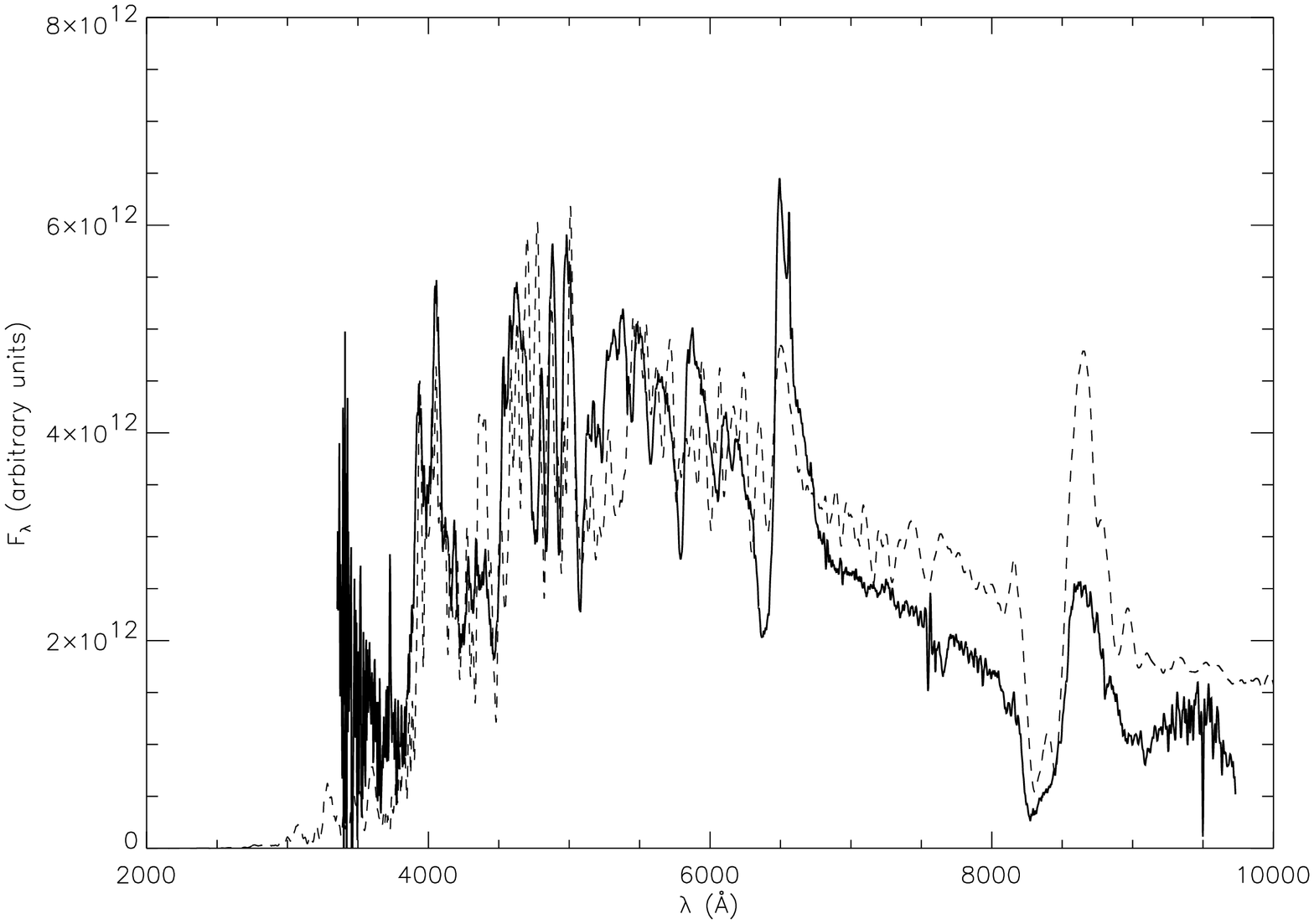}
\caption{Comparison between the observed spectrum at day 37 and the
\texttt{PHOENIX} synthetic spectrum at day 38 after explosion. The
solid line is the observed spectrum, while the dashed line is the
synthetic one.\label{fig:phx1}} 
\end{figure*}

In order to obtain some physical insight with respect to the blue colour of
\98a, we calculated a synthetic spectrum using a hydrodynamical
model of \87a \cite{blinn87a00} and the generalised stellar
atmosphere code \texttt{PHOENIX} \cite{hbjcam99}. Our procedure was
the same as that described in Mitchell et al.
\shortcite{mitchetal87a01,mitchetal87a02}. Fig.~\ref{fig:phx1} shows the
resulting synthetic spectrum compared to the
observed spectrum of \98a~at day 37. Considering that we have not adjusted any parameters, the fit is
reasonable. The synthetic Balmer lines are too weak and the Ca IR triplet is
too strong, but most of the features match the observed ones. 
However, in general, the synthetic lines are too narrow, again indicating that
\98a\/ is likely more energetic than \87a. We stress that the
``photospheric temperature'', i.e. the electron temperature at the
point in the model where the electron scattering optical depth is one,
is about 6000~K, just as expected for SNe II--P at this epoch \cite{popo92}. The blue colour of the
synthetic spectrum comes from the effects of increased $\gamma$--ray heating (see
Mitchell et~al., 2001). In fact, higher $\gamma$--ray heating is likely
required in order to reproduce the strength of the synthetic H$\alpha$
feature, to reduce the flux in the red the strength of
the Ca IR triplet. All this confirms that \98a\/ ejected more
$^{56}$Ni than \87a.
\section{Discussion}

\begin{figure}
\includegraphics[width=14.5cm]{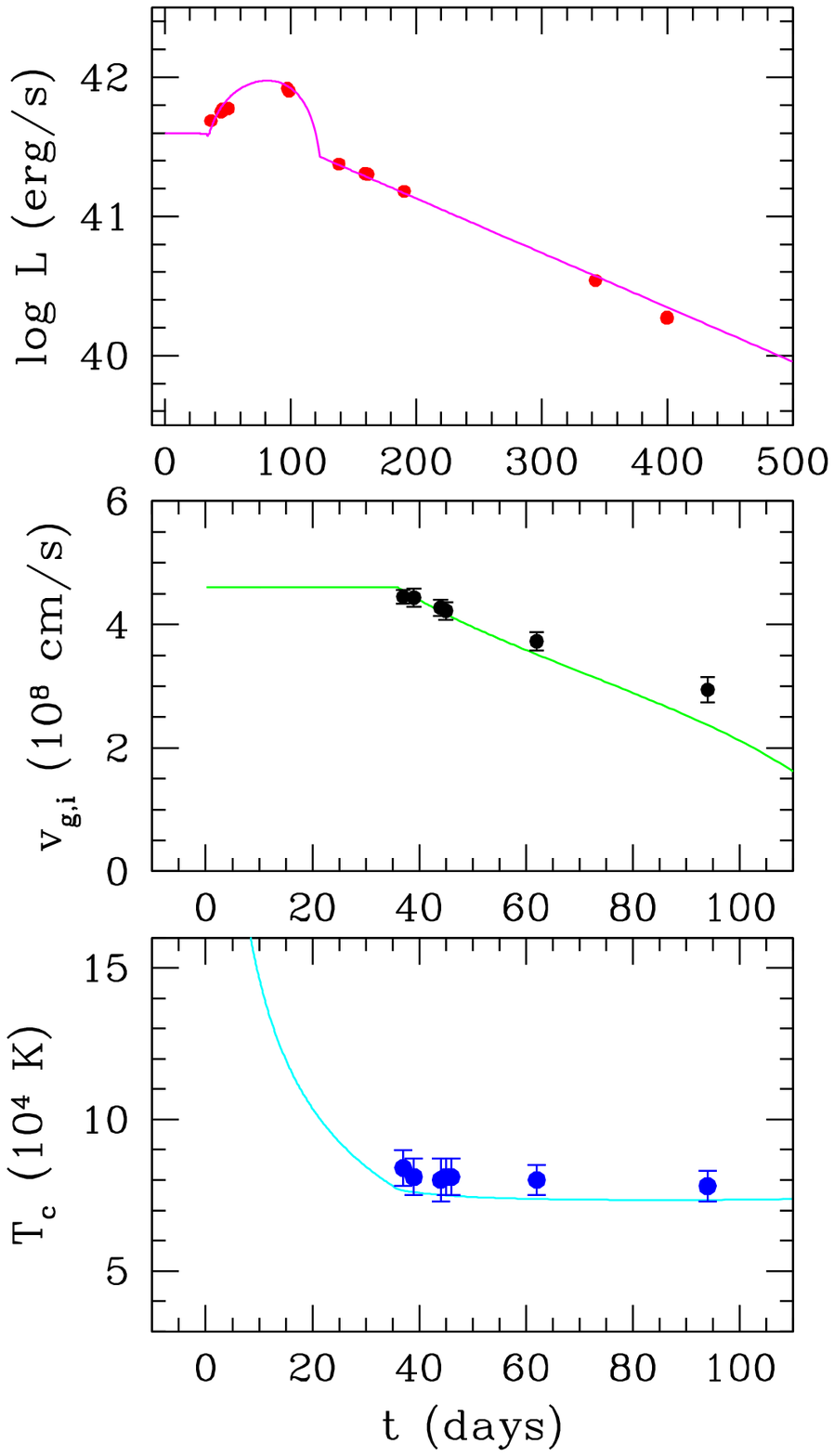}
\caption{Evolution of the luminosity ({\bf Top}), expansion velocity
({\bf Centre}) and continuum temperature ({\bf Bottom}) as functions of time:
comparison between the model (solid lines) and the observations
(filled circles).\label{model}} 
\end{figure}

The observations show that, in spite of the similar evolution of the
light curves and spectra, \98a\/ and \87a\/ have a number of important
differences. Assuming a distance modulus $\mu$ = 32.41 (see
Sect. \ref{lcurv}), \98a\/ has a higher luminosity at all epochs,
likely due to the ejection of a larger amount of $^{56}$Ni. It also
has higher colour temperatures, and higher expansion velocities,
suggesting a more energetic explosion than that of \87a\/.  The low
luminosity at early phases and the presence of a broad peak in the
light curve indicate
that \98a\/ exploded when the progenitor star was in the blue
supergiant stage.  In order to obtain an approximate estimate of the
explosion parameters and gather some information on the nature of the
progenitor star, we have compared the data of SN 1998A with the output
of the semi--analytic code developed by Zampieri et al. (2003). The
model incorporates all the main sources of energy powering a CC--SN
and calculates the evolution of the expanding envelope from the
photospheric phase up to the late nebular phase (with the exception of
shock breakout which is not included). It assumes idealised initial
conditions that provide an approximate description of the ejected
material as derived from hydrodynamical calculations. The parameters
of the ejected envelope are estimated performing a simultaneous
comparison of the observed and computed light curve, photospheric gas
velocity and continuum temperature. The results of the fit for SN
1998A are shown in Fig. \ref{model}, where we assume that the
explosion epoch occurred $\sim$ 19 days before discovery. The main
parameters of the post--shock, ejected envelope are listed in Table
5. The fraction of the initial energy that goes into kinetic energy
$f_0$ and the gas opacity $\kappa$ are input physical constants. In
this calculation we adopt $f_0=0.5$ (initial equipartition between
thermal and kinetic energies) and $\kappa=0.2$ cm$^2$ g$^{-1}$
(appropriate for an envelope comprised of He and iron--group
elements). The colour correction factor $f_c=T_c/T_{eff}$ measures the
deviation of the continuum radiation temperature $T_c$ from the
blackbody effective temperature $T_{eff}$. The value of $f_c$ required
to fit the data of SN 1998A lies in the high end tail of the typical
range for SN II (see e.g. Eastman, Schmidt \& Kirshner, 1996). This is
probably the consequence of the different physical conditions at the
photosphere of SN 1998A.

Despite the similarity in the photometric evolution, a comparison of
the inferred parameters of SN 1998A with those of SN 1987A (also
reported in Table 5) shows that there are significant differences
between the ejecta of these two SNe. In fact, although the initial
radius is small in both objects ($\sim$ 5 $\times$ 10$^{12}$ cm), 
the ejected mass is different 
(18 M$_\odot$ in SN~1987A, 22  M$_\odot$ in SN~1998A) and the total 
energy a few times larger in SN 1998A (5--6 $\times$ 10$^{51}$ erg). 
This is shown by the higher expansion velocity and
continuum temperature. The large inferred envelope mass and total
energy of SN 1998A places it in the upper end of the correlations
among the parameters of the ejecta of SN II recently reported by
Hamuy (2003), Pastorello (2003) and Zampieri et al. (in preparation). This suggests that the
properties of the SN II ejecta vary continuously even in
1987A-like events. The large ejected mass of
SN 1998A implies a fairly massive progenitor, that may form more
easily in a low metallicity environment. 

\begin{table*}
\footnotesize
\caption{Parameters of SN 1987A and SN 1998A from the semi-analytic model.}\label{parameters}
\begin{center}
\begin{tabular}{ccccccccccc}
\hline \hline
SN & $R_0$ & $M_{ej}$ & $M_{Ni}$ & $V_0$ & $E$ & $f_0$ & $\kappa$ & $t_{rec,0}$ & $T_{eff}$ & $f_c$ \\
 & ($10^{12}$ cm) & ($M_\odot$) & ($M_\odot$) & ($10^8$ cm s$^{-1}$) & ($10^{51}$ erg) & & (cm$^2$ g$^{-1}$) & (days) & (K) \\
\hline \hline
1987A  & $\la 5$ & 18 & 0.075 & 2.7 & 1.6 & 0.5 & 0.2 & 25 & 4800 & 1.1 \\
1998A  & $\la 6$ & 22 & 0.11 & 4.6 & 5.6 & 0.5 & 0.2 & 36 & 4500 & 1.6 \\
\hline
\end{tabular}

$R_0$ is the initial radius of the ejected envelope at the onset of
expansion

$M_{ej}$ is the ejected envelope mass

$M_{Ni}$ is the ejected $^{56}$Ni mass

$V_0$ is the velocity of the envelope at the outer shell (i.e. at $R_0$)

$E$ is the initial thermal+kinetic energy of the ejecta

$f_0$ is the fraction of the initial energy that goes into kinetic
energy

$\kappa$ is the gas opacity

$t_{rec,0}$ is the time when the envelope starts to recombine

$T_{eff}$ is the effective temperature during recombination

$f_c=T_c/T_{eff}$ is the colour correction factor

\end{center}
\end{table*}

One of the most intriguing properties of \98a\/ is that the Ba II
lines are fainter than those of \87a\/ and nearly equal to the typical
strength in a normal Type II event [e.g. SN 1988A \cite{tura93} and SN
1996W \cite{pasto0}]. Mazzali $\&$ Chugai (1995) suggest two
possibilities to explain the difference in strength of the Ba II
lines: an intrinsic difference in the abundance of Ba in the ejecta
and different physical conditions. Although it is impossible to
exclude local overproduction of Ba (e.g. a scenario involving binary
systems or stellar mergers, or higher degree of mixing in the envelope
of the Ba synthesised via $s$--process in the He core), the
$s$--process nucleosynthesis in Type II progenitors cannot explain a
Ba overabundance in the envelope larger than 40$\%$. This is well
below that deduced from the strength of the Ba II lines in SN 1987A (a
factor 20; Utrobin $\&$ Chugai, 2002). Therefore, the different
physical conditions at the photosphere are likely to play a crucial
role in determining the appearance or disappearance of strong Ba
lines. In particular, a lower temperature caused by smaller
$\gamma$--ray heating may be a key ingredient.  If in the atmosphere
of most SNe II--P barium is present as Ba III due to non--thermal
ionization by $\gamma$--rays, the observed spectra will show
``normal'' strength Ba II lines. On the contrary, in SN 1987A
(but also in SN 1997D and similar events, Turatto et al., 1998;
Pastorello et al., 2004) lower amounts of $^{56}$Ni ejection produce
less non--thermal ionization and increase the abundance of Ba II ions
\cite{chu88b}.  In addition, we note that the almost normal strength
of the Ba II lines in the spectra of \98a\/ suggests that these lines
are probably not sensitive to the initial radius.

Another observational property of \98a\/ is
the low intrinsic polarization at early phases \cite{leon01}: this indicates
relatively low asphericity of the emitting region, supporting the general scenario
that core--collapse events which retain hydrogen envelopes at the epoch 
of explosion are not significantly aspherical during the
early photospheric phase 
\cite{leon01}.

The  differences between the progenitor properties of \98a\/ and \87a\/
suggest that stars with different masses can explode in the blue
supergiant stage. 
In particular, the discovery of a few SNe II with light curves
similar in shape to SNe 1987A and 1998A [SN 1909A \cite{sand74,bran89}, 
SN 1982F (Yamagata $\&$ Iye, 1982; Tsvetkov, 1984; Tsvetkov, 1988), SN 1998bt \cite{lisa03} 
and SN 2000cb \cite{hamu01}], but showing a significant spread in the magnitudes
at maximum, is an indication that these objects are moderately common, 
even if less frequent than predicted by previous works (e.g. Schmitz
$\&$ Gaskell, 1988). 
In analogy to more typical SNe II--P \cite{hamu01,hamuysnii02,pasto0}, 1987A--like events 
are therefore expected to cover a wide range of physical properties. 

\section{Conclusions}

In this paper we have presented the spectroscopic and photometric data of
the peculiar \98a\/, which shows clear observational analogies
with \87a\/. However, despite the spectro--photometric
similarities, the physical parameters characterising both \87a~and \98a, and
their progenitors, are different.

Optical photometry indicates that the progenitor of \98a\/~was a compact 
blue supergiant and that the progenitor star mass was $\sim$ 25 M$_{\odot}$.

The high line velocities in \98a\/~indicate that the explosion energy
is larger than that of \87a\/: the total initial energy of the ejecta 
soon after the explosion is about 4 times higher than in \87a\/.

The luminosity of \98a\/ exceeds that of \87a\/ at all epochs. 
In particular, the luminosity of the radioactive tail indicates
that the amount of $^{56}$Ni ejected by the explosion of \98a\/~is about 0.11
M$_\odot$ (as compared to 0.075 M$_\odot$ in \87a\/).

In \98a\/ the Ba II P--Cygni lines are fainter than those observed in \87a\/ and
in several other faint CC--SNe \cite{pasto2}, suggesting
that a relation exists between the strength of the Ba II lines and the
ejected $^{56}$Ni mass.

\section*{Acknowledgments}
This paper is based on observations collected at ESO -- La Silla
(Chile).\\
A. Pastorello is grateful to the University of Oklahoma
for generous hospitality during part of this work,
and to D.~Casebeer, D.~Richardson and R.~C.~Thomas
for useful discussions. \\
This work was supported in part by NASA grant
NAG5-12127, NSF grant AST-0204771, and an IBM SUR grant at
the University of Oklahoma.  \\
Some of the calculations presented here were
performed at the San Diego Supercomputer Center (SDSC), supported by
the NSF, and at the National Energy Research Supercomputer Center
(NERSC), supported by the U.S. DOE.  We thank both these institutions
for a generous allocation of computer time.\\
This research has made use of the NASA/IPAC Extragalactic
Database (NED) which is operated by the Jet Propulsion Laboratory,
California Institute of Technology, under contract with the National
Aeronautics and Space Administration. We have also made use
of the Lyon--Meudon Extragalactic Database (LEDA), supplied
by the LEDA team at the Centre de Recherche Astronomique 
de Lyon, Observatorie de Lyon.

\end{document}